\documentclass[twocolumn,amssymb,showpacs,aps,prl]{revtex4-1}
\usepackage{}
\usepackage{stmaryrd}
\usepackage{amsmath}
\usepackage{amssymb}
\usepackage{float}
\usepackage{graphicx}
\usepackage{titletoc}
\usepackage{booktabs}
\usepackage{array, color}
\usepackage{tabularx}
\usepackage{latexsym,bm}
\usepackage{graphics}

\begin{document}

\title{Sub-cycle coherent control of ionic dynamics via transient ionization injection}

\author{Qian Zhang$^{1,\dag}$}
\author{Hongqiang Xie$^{1,2,\dag}$}
\author{Guihua Li$^{1,3}$}
\author{Xiaowei Wang$^{1}$}
\author{Hongbin Lei$^{1}$}
\author{Jing Zhao$^{1}$}
\author{Zhiming Chen$^{2}$}
\author{Jinping Yao$^{4}$}
\author{Ya Cheng$^{5}$}
\author{Zengxiu Zhao$^{1}$}
\email{zhao.zengxiu@gmail.com}
\affiliation{$^1$Department of Physics, National University of Defense Technology, Changsha 410073, China\\
$^2$State Key Laboratory of Nuclear Resources and Environment, East China University of Technology, Nanchang 330013, China\\
$^3$School of Science, East China Jiaotong University, Nanchang 330013, China\\
$^4$State Key Laboratory of High Field Laser Physics, Shanghai Institute of Optics and Fine Mechanics, Chinese Academy of Sciences, Shanghai 201800, China.\\
$^5$State Key Laboratory of Precision Spectroscopy, East China Normal University, Shanghai 200062, China\\
$\dag$These authors contributed equally to this work.}

\date{\today}

\begin{abstract}
We investigate the interwoven dynamic evolutions of neutral nitrogen molecules together with nitrogen ions created through transient tunnel ionization in an intense laser field. By treating the molecules as open quantum systems, it is found that considering real-time injection of ions and strong couplings among their electronic states, nitrogen molecular ions are primarily populated in the electronically excited states, rather than staying in the ground state  as predicted by the well-known tunneling theory. The unexpected result is attributed to sub-cycle switch-on of time-dependent polarization by transient ionization and dynamic Stark shift mediated near-resonant multiphoton transitions. Their combined contribution also causes that the vibrational distribution of N$_2^+$ does not comply with Franck-Condon principle. These findings corroborate the mechanism of nitrogen molecular ion lasing and are likely to be universal. The present work opens a new route to explore the important role of transient ionization injection in strong-field induced non-equilibrium dynamics. 
\end{abstract}


\maketitle

The fundamental process of strong field ionization of atoms by intense ultrashort laser pulses occurs at attosecond timescale \cite{Eckle08s,Schultze2010,Klunder11} and lasts for pulse lengths in femtoseconds. The temporal confined strong field ionization  (SFI) creates broad bandwidth non-stationary ionic states  along with launching of attosecond electron wavepacket  forming the foundation of attoseond physics \cite{Goulielmakis2010,Wirth2011,Worner2011,Ossiander2017,Sabbar2017}. One of the key problem is how the coherence of the ionic states affect the subsequent ultrafast non-equilibrium  evolution  ranging from charge migration in molecules \cite{Cederbaum99}, electron transport  in condense matter \cite{Cavalieri07} to ultrafast laser processing of materials \cite{Malinauskas2016}.   It is particular intriguing to question wether  the further interaction of the ions with lasers can be  considered independently by assuming the prior ionization is completed? 


Recent experiments  indicate that  the interplay of  sub-cycle SFI and the followed ion-laser coupling is indispensable for nitrogen molecular ion lasing \cite{Xu2015,Yao2016,LiuYi2015,LiHL2019,Mysyrowicz} which suggests the new possibility to  manipulate the ion coherence upon its creation toward ion-based quantum optics. While SFI itself can be described well by the celebrated Keldysh tunneling theory \cite{Keldysh65}, a complete model treating both ionization of neutrals and laser-ion couplings on the equal footing is still lacking.  Theoretically, dealing with bounded multi-electron problem is already a difficult task, it is even challenging for open quantum many-body systems when ionization is involved. Exact time-dependent multilelectron theories are limited to two electrons cases \cite{Hart07} or struggling with ionization-induced derivative discontinuities  in density-functionals \cite{ Lein05L1}. A lot of theoretical works have devoted to the coherent ionic evolution in a multi-channel formalism \cite{Santra06a,Rohringer2009} but the ion-laser coupling {\bf within} the ionization process remains elusive.

\begin{figure}
\includegraphics*[width=3in]{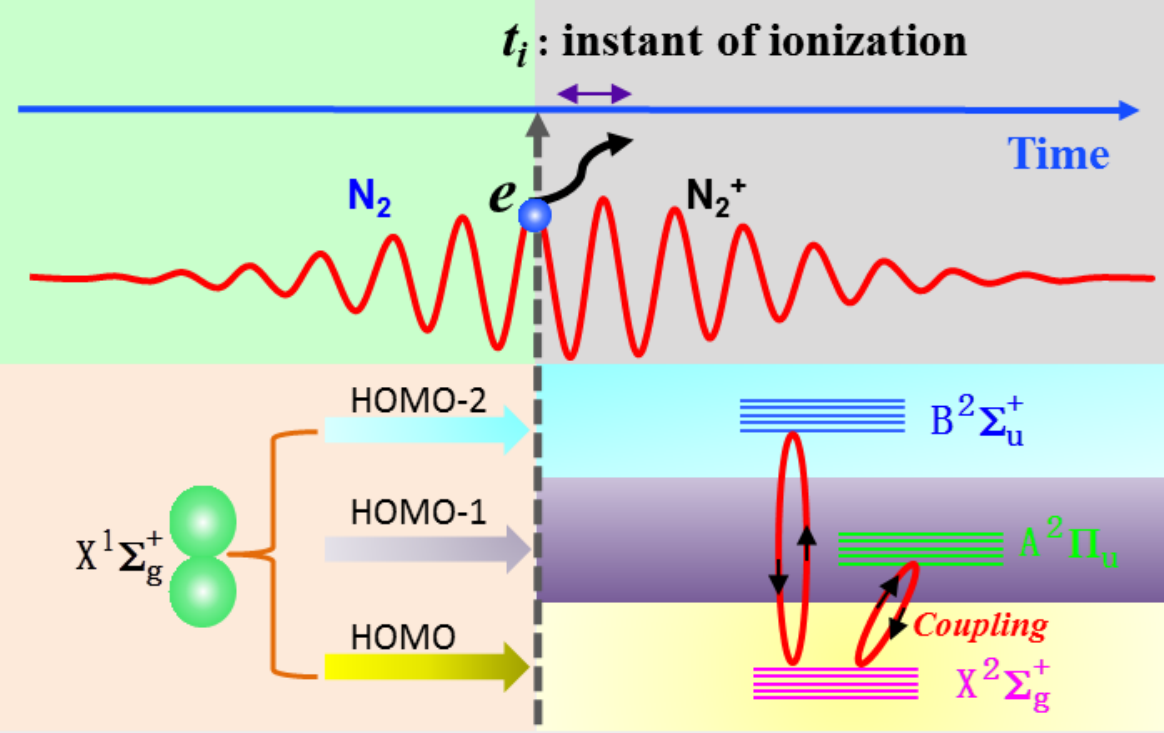}
\caption{(Color online)
Illustration of the dynamic processes of nitrogen molecules in an intense laser pulse. Strong field ionization injects  the ions into three possible electronic states that are driven by the remaining laser pulse.  Note that the X-B transition is parallel and  X-A transition is perpendicular to the molecular axis.
}\label{f1}
\end{figure}

The current work attempts to address the fundamental question of how the transient SFI  will influence the quantum states of the residual ions and their coherent evolution in the laser fields. By treating the nitrogen molecule as an open quantum system,  we systematically investigate the role  of transient injection of ions by SFI on ultrafast population redistribution of ionic states  based on a new theoretical frame. As illustrated in Fig.~\ref{f1}, ionization of a nitrogen molecule creates the ion in three possible electronic states  at the moment $t_i$ by releasing the electron, the remaining linearly polarized laser field will further induce coherence and population transfer among the electronic and vibrational states with the couplings depending on the geometry of the molecular ion. 

According to the  tunneling ionization theory \cite{Keldysh65,Tong02}, the population on the excited ionic states is normally much less than that on the ionic ground state.  Due to the continuous presence of the laser field, the ions will be polarized causing the electron to oscillate back and forth between electronic states.  As we will demonstrate, the sub-cycle turn-on of the polarization of the ion upon its creation breaks the time-reversal symmetry and it is thus possible to enhance the population on the higher excited states. Furthermore, in the case of resonance, the population on the excited states can also be greatly increased by multiphoton couplings mediated by dynamic Stark shift due to the instantaneous polarization.  The ultrafast population redistribution of electronic states of molecular ions assisted by transient ionization might be a universal process for other molecular ionization from multiple orbitals and bond breaking and thus SFI can serve as an intrinsic attoseond probe to the sub-cycle ionic coherence. Our finding sheds new light on the strong field ionization-coupling mechanism and provides crucial implications for further research on coherent emissions from molecular ions.

{\bf Theory.} We consider tunnel ionization mainly from three N$_2$ molecular orbitals, i.e., HOMO, HOMO-1, and HOMO-2 in a linearly polarized pump laser field with the peak intensity of $3\times 10^{14}$ W/cm$^2$, pulse duration of $15$ fs and the central wavelength of $800$ nm. The ionization energies of these orbitals are respectively 15.6 eV, 17 eV and 18.8 eV \cite{Chong2002}. MO-ADK theory is employed to calculate the transient ionization rates of the involved three orbitals \cite{Tong02}. Once the three ionic states   $X^2\Sigma_g^+$ (i.e., $X$),  $A^2\pi_u$ (i.e., $A$) and  $B^2\Sigma_u^+$ (i.e., $B$) are prepared at the moment of ionization, population couplings among them will take place in the residual laser field. 

Considering both the transient ionization injection and coupling effects, the  evolution of the ionic density matrix $\rho^+(t)$  in an intense laser field is given by
\begin{eqnarray}
\frac{d(\rho^+)}{dt}=-\frac{i}{\hbar}\lbrack H_I(t),\rho^+(t)\rbrack+\left(\frac{d(\rho^+)}{dt}\right)_{ionization}
\end{eqnarray}
where $H_I$ is the ionic Hamiltonian in the interaction picture whose explicit form is provided in  \cite{Supp}.  The last term  in Eq.~(1) is one of the significant advances in this work.  It represents the continuous injection of ions  by transient ionization within the full laser pulse and  is given by
\begin{eqnarray}
\left(\frac{d(\rho^+_{iv)}}{dt}\right)_{ionization}=\Gamma_{iv}(t)\rho_0
\end{eqnarray}
where  $i= X, A, B$ labels the electronic states and $v=0\sim4$ are the vibrational quantum numbers. 
The neutral population $\rho_0$ is decaying according to $\frac{d\rho_0}{dt}=-\sum \Gamma_{iv}\rho_0$ where $\Gamma_{iv}(t)$ are the ionization rates to the respective electronic-vibrational states. For the time being, the rotational states are unresolved and dissipation by collisions are ignored because of  the short pulse duration. 

\begin{figure}
\includegraphics*[width=3in]{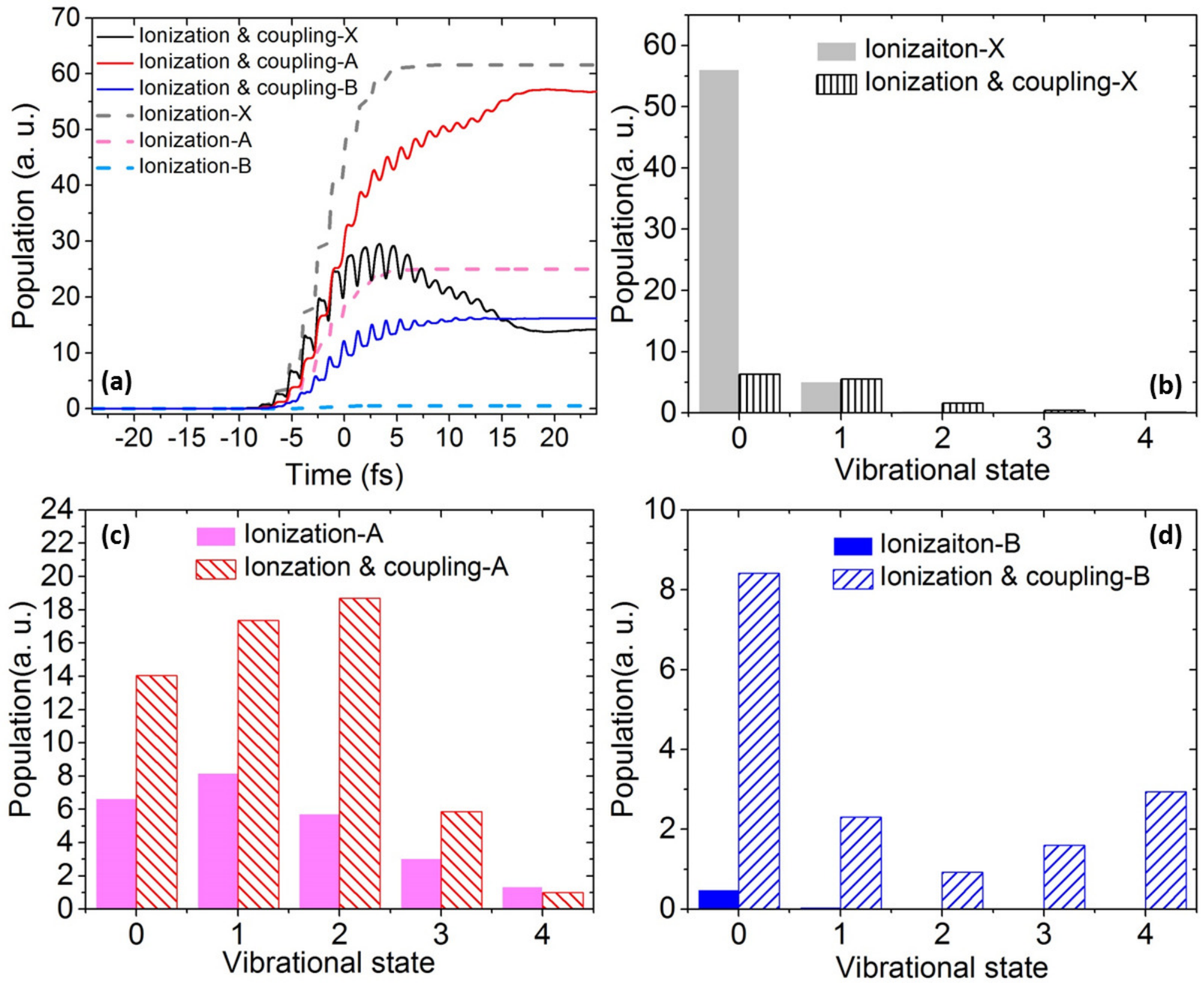}
\caption{(Color online)
(a) Dynamic evolution of electronic state population  of N$_2^+$ taking into account of both  ionization and laser-ion coupling (solid curves), and that obtained by considering only the ionization (dashed curves). The molecular axis is aligned to be $45^o$ with respect to the laser polarization direction. The final vibrational state-resolved population distribution for each electronic state after interaction are respectively shown in (b), (c) and (d). 
}\label{f2}
\end{figure}

{\bf Population inversion and vibrational redistribution.}  
We now present our results and discussions.  The details of calculation can be found in \cite{Supp}.
We first discuss the nitrogen molecule whose axis is aligned to be 45$^o$ with respect to the pump laser polarization and therefore both the transitions of X-A, X-B are permitted. The rotational degree of freedom is assumed to be frozen as mentioned previously. 
The solid curves in Fig.~\ref{f2}(a) shows the population evolution of the three electronic states  of N$_2^+$ in the pump laser field by solving ionization-coupling equation. It can be clearly seen that the final populations on the states $A$ and $B$ are prominently increased while that on the state $X$ is greatly decreased in comparison to the case that only ionization is considered (the dashed curves in Fig.~\ref{f2}(a)). It can be seen that with the help of the transient injection the population between the states B and X is thus inverted. This is striking in view of that  other processes, e.g, shake-up \cite{Litvinyuk05,Bryan06} or recollision \cite{Britton2018,LiHX2019} are unlikely to produce ionic population inversion. 

The vibrational state-resolved distributions for each electronic state  are respectively displayed in Fig.~\ref{f2}(b), (c) and (d). As can be seen, for the electronic state $X$, the ionization itself mainly populates the vibrational  states $v=0, 1$ because of the relatively larger Frank-Condon factors of the two states \cite{Alf1977}. However, when the coupling is incorporated, the vibrational population on the state $v=0$ is largely reduced while higher vibrational states are more populated. It can be attributed to  vibrational Raman-like processes when a strong coherent coupling is produced between the states $X$ and $A$.  As a consequence of the $X$ state population reduction, almost all the vibrational population on the state $A$ are efficiently enhanced due to one-photon X-A resonant transition. Remarkably, the vibrational-state population on the state $B$ (especially for $v=0$) are considerably increased  as well although the laser frequency is far off from X-B resonant energy. Note that the vibrational energy gap between the states $A$($v=0\sim4$) and $X$($v=0$) spans from 1.12 eV to 2.06 eV and the driving laser photon energy ranges from 1.38 eV to 1.77 eV (700-900 nm). For the excited state $B$($v=0\sim4$), the vibrational energy differences with respect to the ground state $X$($v=0$) are in the range of 3.17 eV$\sim$4.37 eV. The population transfer from the state $X$($v=0$) is only accessible to high vibrational states $B$($v=3, 4$)  through a direct three-photon coupling.

\begin{figure}
\includegraphics*[width=2.9in]{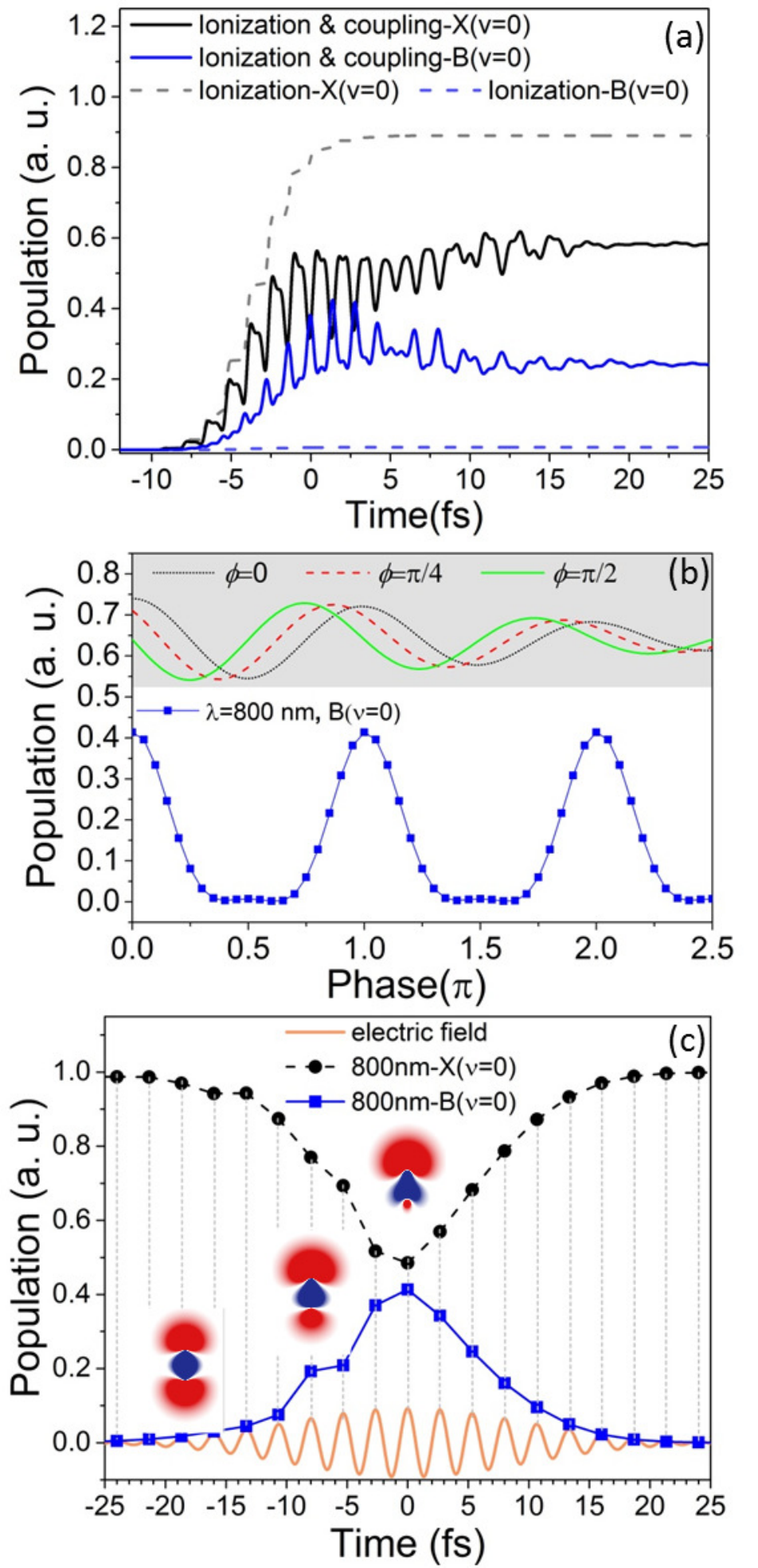}
\caption{(Color online)
(a) Dynamic evolution of electronic state populations of N$_2^+$ when the molecular axis is parallel to the laser polarization. (b) variation of $B$($v=0$) population  with the initial injection phase $\phi$
after interacting with  a half laser pulse at 800 nm. The corresponding laser fields for $\phi=0,\pi/4,\pi/2$ are respectively depicted. (c) dependence of the yield population on the injection instants within the full laser pulse. At these instants, the laser field takes local maximum at each individual optical cycle.
The contour plot illustrates the transient ionic states that are polarized according to the instantaneous laser field.
}\label{f3}
\end{figure}

{\bf Sub-cycle control of laser-ion coupling via transient ionization-injection.}  In order to better understand the observed population enhancement of the lower vibrational states (i.e., $v=0, 1, 2$) of the state $B$, we fixed nitrogen molecular axis to be parallel to the laser polarization, by which the A-X coupling is avoided. In the following, we mainly focus on the $B$ ($v=0$) state  because the manipulation of its population is critical to air lasing at 391 nm \cite{Yao2011,XieHQ2014,LiuYi2015, Zhong2017,LiHL2019, Xu2015,MiaoZM2018}

Figure 3(a) shows the evolution of the vibrational population of the ionic states $B$($v=0$) and $X$($v=0$) for the parallel alignment case. Again, the population on the state $B$($v=0$) is apparently promoted with the aid of simultaneous ionization and coupling. To investigate the influence brought by ionization on the coupling between $B$($v=0$) and $X$($v=0$), we first consider the ionization injection at the peak laser intensity and then follow the evolution of the ions in the second half of the laser field $E(t)=f(t)\cos(\omega t+\phi)$ with different phase $\phi$. The corresponding laser fields for $\phi=0, \pi/4, \pi/2$ are respectively depicted in the top portion of the Fig. 3(b). The injection is all populated on the ground state $X$ of N$_2^+$, i.e., $\rho^+_{XX}(v=0)=1$.  
Fig.~\ref{f3}(b) plots the final population of the state $B$($v=0$) as a function of the initial phase $\phi$ for the pump wavelength 800 nm. Interestingly, the obtained population on the state $B$($v=0$) is strongly modulated with a period of $\pi$ and the maximum value is achieved at $\phi=0$. This explains that the yield population  calculated in \cite{Yao2016} who chose $\phi=\pi/2$ is less than that calculated in \cite{Xu2015}  who chose $\phi=0$ using their three-state coupling model. 

We now consider how the coupling varies with the field profile assuming the ionization injection occurs at the peak field of each optical cycle (corresponding to fixed $\phi=0$). The population on the $B$($v=0$) and $X$($v=0$) states of N$_2^+$ after the laser-ion coupling are shown in Fig.~\ref{f3}(c) as a function of the instant of transient ionization injection. The yield population on the state $B$($v=0$) shows a nearly Gaussian dependence on the moments of ionization injection following the intensity profile of the pumping laser, as depicted by the blue squares Fig.~\ref{f3}(c). The stronger of transient electric fields at the injection of the ground-state ions, the more population on the state $B$($v=0$) and reversely the less population on the state $X$($v=0$) are found after the pulse is finished.

{\bf  Sub-cycle switch-on  and probing of ionic polarization.}  The above observations can be qualitatively explained with ultrafast polarization theory. At a particular time, nitrogen molecular ions are prepared by transient SFI, which can be regarded as an ultrafast pump for the ionic system. Immediately following SFI, the ion is polarized by the  instantaneous laser field as illustrated by the contour plot shown in Fig. 3(c). 

Since the electronic states $B$ and $X$ forming a pair of charge resonance states, strong polarization could occur due to their coherent coupling  in the residual laser field. For different injection instants of ionization, the polarization state of ions is different, resulting in different population increment of the state $B$($v=0$). For the case in Fig. 3(b), the probability for populating on the state $B$($v=0$) is greatest at $\phi=0$ while at $\phi=\pi/2$, the probability is smallest. For comparisons of different ionization injection moments at the same envelope phase in Fig. 3(c), the final population on the state $B$($v=0$) varies because the instantaneous polarizability is proportional to the transient electric field. Therefore, transient ionization injection acts as an ultrafast probe of a quantum coherent system consisting of the neutral and molecular ions. 

\begin{figure}
\includegraphics*[width=3in]{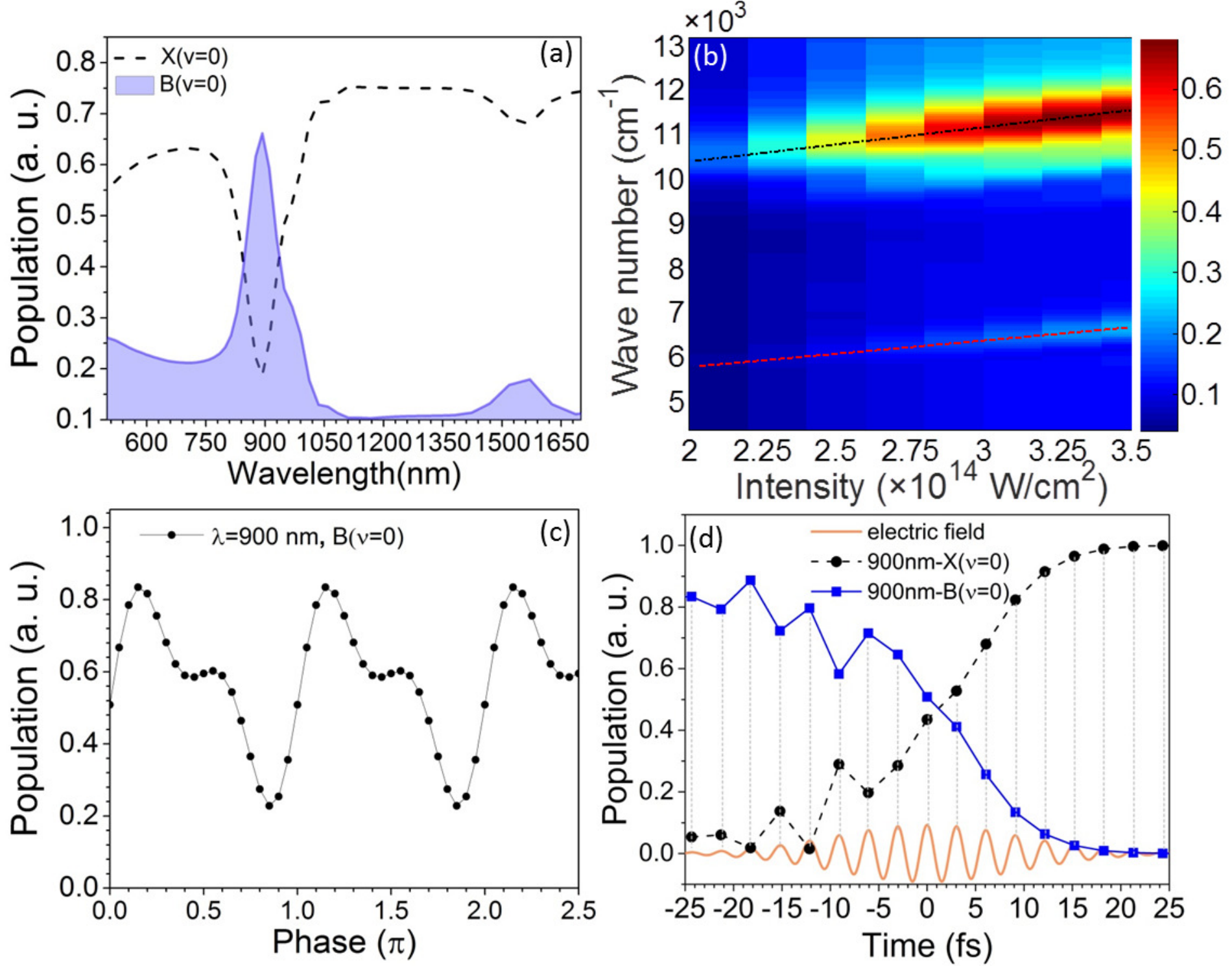}
\caption{(Color online)
(a) final population on the state $B$($v=0$) and $X$($v=0$) as a function of the central laser frequency. (b) contour plot of the final population verse laser intensity and frequency.  The  three-photon and five-photon resonant frequencies shift with  the applied laser peak intensity is indicated by the corresponding linear fittings (black dash-dotted and red dashed curves). (c) the dependence of  $B$($v=0$) population on the initial injection phase $\phi$ at  the central wavelength of 900 nm. (d) dependence of $B$($v=0$) and $X$($v=0$) populations on the moments of ionization injection by fixing the carrier phase in the near resonant case at the central wavelength of 900 nm.
}\label{f4}
\end{figure}
 
{\bf Multiphoton transition mediated by dynamics Stark shift.}  Further enhancement of the population on the state $B$($v=0$) can be achieved by resonant transitions from $X$($v=0$). Since this process is sensitive to the central wavelength of laser pulses, we calculated the population dependence  on the driving laser frequency by solving ionization-coupling equation. As shown in Fig.~\ref{f4}(a), a three-photon resonant peak for $B$($v=0$) appears at the wavelength of ~900 nm, which deviates from the field-free three-photon resonant wavelength (1173 nm). Additional resonant peak around 1560 nm belongs to five-photon resonance. Fig.~\ref{f4}(b) shows the corresponding frequencies of the two resonant peaks as a function of the peak intensity of the driving laser field. Both three photon and five photon resonant peaks exhibit a linear dependence on the peak intensity of the laser field. Surprisingly, the shift of the resonance is independent of the driving frequency which can be attributed to the Stark shift by instantaneous polarization of the laser coupled $X$ and $B$ state at the instant of ionization injection. 

Figure 4(c) shows the sub-cycle control of the population on the state $B$($v=0$) by changing the envelope phase  $\phi$ at the resonant wavelength of 900 nm. It can be seen that the population on the state $B$($v=0$) is also modulated with a period of $\pi$. However, the maximum value is no longer at $\phi=0$ and the minimum value is not reached at $\phi=\pi/2$, which signifies that the polarization mechanism is not dominating in the case of resonance. The irregular modulation (black circles) indicates that a multi-channel interference might contribute to the population increment of the state $B$($v=0$). Fig.~\ref{f4}(d) shows the dependence of the population yields on the injection time.  A striking difference with the results in Fig.~\ref{f3}(c) is that in the case of resonance, the population on the state $B$($v=0$) is closely related to the post-ionization interaction time. The population on the state $B$($v=0$) shows a gradual decay with decreasing coupling time in the current conditions. It should be noted that the three-photon resonant transfer from $X$($v=0$) to $B$($v=0$) at a resonant wavelength is taking place during the evolution of the quantum coherent system driven by ultrafast polarization. Therefore the population increment of the state $B$($v=0$) should originate from the two-channel interference, i.e., polarization and three-photon resonant coupling.

{\bf  Attosecond streaking by SFI.}  The injection of ionization can be considered as an ultrafast streaking of the laser driven ionic dynamics analogy to the attosecond streaking and transient absorption techniques \cite{Itatani02,Goulielmakis2010,Wirth2011,Chini2012,Wirth2011}. The Fourier transformation of the phase-dependent populations on the state $B$ in Fig. \ref{f4}(c) quantitatively gives the relative contributions of multiple photon dressing. A concrete analysis on this based on an analytic solution is given in \cite{Supp}. It is worth mentioning that for the non-parallel alignment case, transient ionization could influence both the coupling of A-X and B-X. Last but not the least, in the current model, we have ignored the ionization delays while injection of ions and the coherence brought by ionization. A more complete calculation including these effects will be carried out in the future.

In conclusion, we have systematically investigated the coherent evolution of nitrogen molecular  ions, which are continuously injected by transient strong field ionization in an intense laser field. It is found that the population on the excited states of nitrogen molecular ion can be greatly increased due to transient-injection induced collapse of a quantum system and resonant multiphoton couplings. Our findings provide crucial clues to create high intensity air lasers using femtosecond laser pulses and highlight the importance of ionization induced coherence. The proposed new theoretical frame allows us to treat transient ionization and laser-ion coupling for open quantum systems under strong laser fields. It can be generalized to explore the important role of transient ionization injection in other strong field induced non-equilibrium dynamics, e.g., localization of electrons during dissociative ionization and autoionization of molecules in intense laser pulses. 

\begin{acknowledgments}
This work is supported by the Major Research plan of NSF (Grant No. 91850201), the National Natural Science Foundation of China (Grant Nos. 61705034, 61605227 and 11704066) and Natural Science Foundation of Jiangxi Province (Grant No.20171BAB211007), Science and Technology Project of Jiangxi Provincial Education Department (Grant No.GJJ160587 and GJJ160576).
\end{acknowledgments}


%

\end{document}